\documentclass[11pt]{article}
\usepackage{amsmath,amssymb,color,graphics,epsfig,cite,CJK}
\usepackage{url}
\usepackage[colorlinks=true,citecolor=blue,linkcolor=black]{hyperref}
\usepackage{amsfonts}
\newcommand{\be}{\begin{equation}}
	\newcommand{\ee}{\end{equation}}
\newcommand{\bea}{\setlength\arraycolsep{2pt} \begin{eqnarray}}
	\newcommand{\eea}{\end{eqnarray}}

\def\0{{\sst{(0)}}}
\def\1{{\sst{(1)}}}
\def\2{{\sst{(2)}}}
\def\3{{\sst{(3)}}}
\def\4{{\sst{(4)}}}
\def\5{{\sst{(5)}}}
\def\6{{\sst{(6)}}}
\def\7{{\sst{(7)}}}
\def\8{{\sst{(8)}}}
\def\sst#1{{\scriptscriptstyle #1}}

\def\M2{{\bar{\mathcal{M}}}_{(2)}}
\def\Mco2{{\hat{\mathcal{M}}}_{(D-2)}}
\begin{document}
	\begin{CJK}{UTF8}{song}
		\begin{center}
				{\Large {\bf %Revisit
				A trick for calculating surface gravities of Killing horizons
			}}
			
			\vspace{40pt}
			{\bf Jinbo Yang%\hoch{$\dagger$}
			}
			
			\vspace{8pt}
			
			{\it Department of Astronomy, School of Physics and Materials Science,\\
				Guangzhou University, Guangzhou 510006, P.R.China}\\
			
			\vspace{40pt}
			
			\underline{ABSTRACT}
		\end{center}
		We propose a trick for calculating the surface gravity of the Killing horizon, especially for cases of rotating black holes. By choosing nice slices, the surface gravity and angular momentums can be directly read from relevant components of the inverse metric. We give several cases to show how to apply the trick step by step.
	\vfill {\footnotesize yangjinbo@gzhu.edu.cn }
%	yangjinbophy@gmail.com    e-mail changed; Contents about Kodama vector removed; trandictional solving details added; DC conductivity ; More cases
		\thispagestyle{empty}
		
		\pagebreak
		\tableofcontents
		\addtocontents{toc}{\protect\setcounter{tocdepth}{2}}
	
		%%%%%%%%%%%%%%%%%%%%%%%%%%%%%%%%%%%%%%%%
		
		\newpage	
		
		%%%%%%%%%%%%%%%%%%%%%%%%%%%%%%%%%%%%%%%%

%%%%%%%%%%%%%%%%%%%%%%%%%%%%%%%%%%%%%
\section{Introduction}
Black holes, as the robust prediction of general relativity \cite{Penrose:1964wq}, their existence is strongly supported by  astronomical observations\cite{EventHorizonTelescope:2019dse, EventHorizonTelescope:2019ths, EventHorizonTelescope:2022wkp, EventHorizonTelescope:2022wok, LIGOScientific:2016lio, LIGOScientific:2016sjg, LIGOScientific:2017bnn}. 
A black hole is characterized by its event horizon. For a stationary black hole, its event horizon coincides with the so-called Killing horizon, the null hypersurface where the particular Killing vector field becomes its null normal vector\cite{WaldGR, LiangGRvIII}. The surface gravity $\kappa$ of the Killing horizon plays an important role in black hole thermodynamics. Hawking applied the quantum field theory in curved spacetime to predict a black hole emits thermal radiations that have temperature proportional to the surface gravity $\kappa$ \cite{Hawking:1974rv, Hawking:1975vcx}. A non-zero $\kappa$ indicates that the black hole is not extreme. Thus it corresponds to a finite-temperature system. 
In addition, a black hole may finally lose its mass and disappear. Such a prediction leads to the famous black hole information paradox\cite{Hawking:2005kf}.

The idea of nice slices was proposed in Ref.\cite{Mathur:2009hf} which aims to clarify the crucial question in the black hole information paradox. 
Despite other conditions to define nice slices, such an idea at least requires that the set of nice slices should be smooth near the horizon\cite{Mathur:2009hf, Polchinski:2016hrw}. 
Another approach for solving the information paradox, the Kraus-Parikh-Wilczek tunneling formalism for Hawking radiation, uses the Painlev\'e-Gullstrand(PG) time coordinate to give an evolving picture for shrinking horizon\cite{Kraus:1994zu, Kraus:1994by, Kraus:1994fj, Parikh:1999mf, Parikh:2004ih, Parikh:2004rh, Hemming:2000as, Medved:2001ca, Medved:2002zj, Zhang:2005xt, Zhang:2005uh, Zhang:2005wn, Zhang:2005sf, Jiang:2005ba}. The PG time is regular on the horizon. It labels the hypersurfaces regularly penetrating the horizon. Hence those hypersurfaces are nice slices in the modest sense.
Recently, research concerning PG time slices has given alternative viewpoints for studying cosmology and embedding a black hole into an expanding universe\cite{Gaur:2022hap, Volovik:2022cay}.
Nevertheless, we are inspired by the wide-range usage of nice slices to propose the trick for calculating surface gravities.

This article is organized as follows. Section 2 will prove the general theorem hinting at the trick for calculating Killing horizon surface gravity and angular velocities. There is no need to specify a concrete metric, but only to require the existence of Killing horizons and some Killing vector fields. In section 3, we take spherically symmetric spacetime, the three-function generalized Kerr metric given in Ref.\cite{Baines:2023dhq} and the four-function generalized Kerr-de Sitter(dS) metric as examples to show details of applying the trick. Such a treatment directly derives the simple and remarkable result for the surface gravity. Finally, we summarize in section 4.

\section{The theorem}
\subsection{Set up}
We discuss a spacetime containing a set of Killing vector fields  $\{t^{\mu}, \psi_{i}^{\mu}\}$ which communicate with each other.
Namely,  \begin{equation}
	\mathcal{L}_{t}\psi_{i}^{\mu} = \mathcal{L}_{\psi_{i}}t^{\mu}=\mathcal{L}_{\psi_{j}}\psi_{i}^{\mu}=0 \,,
\end{equation}
where the vector field $t^{\mu}$ is the Killing vector field for defining energy while the fields $\psi_{i}^{\mu}$ respond for several angular momentums. We label the number of $\psi_{i}^{\mu}$ as $p$ while the spacetime dimension as $n$.

Another condition is that the spacetime has at least one non-degenerate Killing horizon which its null normal vector field $\chi^{\mu}$ is a particular linear combination of $\{t^{\mu}, \psi_{i}^{\mu}\}$. Concretely, it is
\begin{equation}
	\chi^{\mu} = t^{\mu} + \Omega_H^{i} \psi_{i}^{\mu} \,,\label{nurmalcovec}
\end{equation}
in which $\Omega_{H}^{i}$ are identified as angular velocities in general cases.
The non-degeneracy means the Killing horizon should contain a non-vanishing surface gravity. According to Refs.\cite{WaldGR, LiangGRvIII}, that is
\begin{equation}
	(\chi^{\nu}\nabla_{\nu}\chi^{\mu})_H= \kappa (\chi^{\mu})_H \,.\label{surfacefravitydef}
\end{equation}
in which the subscript $H$ means taking value \textit{on the horizon}. 
These are all conditions that spacetime should satisfy for our purpose: to show a well-selected coordinates frame leads to a simple formula for calculating $\kappa$.

\subsection{Applying the Frobenius theorem}
The conditions seem not enough to derive a concrete formula. To handle it, we will apply the Frobenius theorem which ensures the existence of $n-p-1$ scalars $\{y^A\}$ called Frobenius scalars (see Refs.\cite{WaldGR, LiangGR}). They are potentially some coordinates to specify a spacetime point. 
Frobenius scalars $\{y^A\}$ are charaterized by vanishing results of the Lie-derivative by $\{t^{\mu}, \psi_{i}^{\mu}\}$, i.e., $\mathcal{L}_{t}y^A=\mathcal{L}_{\psi_{i}}y^A=0$.
Fixing $\{y^A\}$ would pick up a so-called Frobenius integral sub-manifold with dimension $p+1$. Thus we still need $p+1$ coordinates to specify a spacetime point.
We choose another set of scalars $\{ v,\varphi^{i} \}$ which satisfy 
\begin{equation}
	\mathcal{L}_{t}v=1\,,\quad \mathcal{L}_{\psi_{i}}v=\mathcal{L}_{t}\varphi^{i}=0\,,\quad \mathcal{L}_{\psi_{i}}\varphi^{j}=\delta^j_i \,. \label{KillingVars}
\end{equation}
We then treat scalars $\{ v,\varphi^{i} \}$ as coordinates for Frobenius integral sub-manifolds.
Combining with Frobenius scalars $\{y^A\}$, we have picked up a coordinates frame a coordinates frame $\{ v,\varphi^{i}, y^A \}$. The Killing fields under this coordinates frame serve as coordinate basis vectors, namely,
\begin{equation}
	t^{\mu}\frac{\partial}{\partial x^{\mu}}= \frac{\partial}{\partial v} \,,\quad 
	\psi_{i}^{\mu}\frac{\partial}{\partial x^{\mu}}= \frac{\partial}{\partial \varphi^i}  \,.
	\label{KillingfinCoor}
\end{equation}
For instance, components $g_{vv}$, $g_{vi}$, and $g_{ij}$ are
\begin{equation}
	g_{vv} = g_{\mu\nu} t^{\mu} t^{\nu} \,,\quad
	g_{vi} = g_{\mu\nu} t^{\mu} t\psi_{i}^{\nu} \,,\quad
	g_{ij} = g_{\mu\nu} \psi_{i}^{\mu} \psi_{j}^{\nu} \,.
\end{equation}
Moreover, the Killing equations $\mathcal{L}_{t}g_{\mu\nu}=\mathcal{L}_{\psi_{i}}g_{\mu\nu}=0$ indicates that $\partial g_{\mu\nu}/\partial v=\partial g_{\mu\nu}/\partial \varphi^i=0$, i.e., all components $g_{\mu\nu}$ should only depend on $y^A$.
As for components $g^{\mu\nu}$, they can be treated as
\begin{align}
	g^{vv}&=g^{\mu\nu} \nabla_{\mu}v\nabla_{\nu}v\,,\;\;\quad 
	g^{vi}=g^{\mu\nu} \nabla_{\mu}v\nabla_{\nu}\varphi^{i}\,,\;\;\quad 
	g^{ij}=g^{\mu\nu} \nabla_{\mu}\varphi^{i}\nabla_{\nu}\varphi^{j} \,, \nonumber
	\\
	g^{vA}&=g^{\mu\nu} \nabla_{\mu}v\nabla_{\nu}y^A\,,\quad 
	g^{iA}=g^{\mu\nu} \nabla_{\mu}\varphi^{i}\nabla_{\nu}y^A\,,\quad 
	g^{AB}=g^{\mu\nu} \nabla_{\mu}y^A\nabla_{\nu}y^B \,.
\end{align}
On the other hand, they serve as the inverse of $g_{\mu\nu}$ such that they also do not rely on $v$ and $\varphi^i$.
We further require $\{ v,\varphi^{i} \}$ \textit{regular} near the horizon, which means that $g^{vv}$, $g^{vi}$, $g^{ij}$, $g^{vA}$, amd $g^{iA}$ are smooth functions in the neibergh of the horizon. Particularly, the coordinate $v$ labels a set of \textit{nice slices} that penetrates the Killing horizon. 
The location of the Killing horizon should be determined by some function $F(y^A)=0$, at least locally. We label $\nabla_{\mu}F\nabla^{\mu}F$ as $g^{FF}$,  $\nabla_{\mu}v\nabla^{\mu}F$ as $g^{vF}$, and $\nabla_{\mu}\varphi^i\nabla^{\mu}F$ as $g^{iF}$ for short. Then we are going to show that the surface gravity $\kappa$ and angular velocities $\Omega^i_H$ can be calculated by
\begin{equation}
	\kappa 	=\frac{1}{2 (g^{vF})_H^2}\,(\nabla^{\mu}g^{FF})_H(\nabla_{\mu}v)_H \,,\quad
	\Omega^i_H = (g^{iF}/g^{vF})_H \,,
	\label{sufgrapritrick}
\end{equation}
in which we still use the subscript $H$ to emphasize taking value \textit{on the horizon}. 
It is worth noting that not all $g^{vA}$ is zero since we have chosen a regular coordinates frame rather than a set of orthonormal coordinates. 
It is due to the properties of $\nabla_{\mu} F$ that we will discuss later.

\subsection{Near the horizon}
Before proofing Eq.\eqref{sufgrapritrick}, we will clarify the equation $F(y^A)=0$ which determines the location of the horizon, and its derivative $(\nabla_{\mu}F)_H=(\partial F/\partial y^A)_H( \nabla_{\mu}y^A)_H$ on the horizon.

Firstly, the location of the Killing horizon is somewhere $\chi^{\mu}\chi_{\mu}$ becoming zero, i.e., $g_{vv}+2g_{iv}\Omega^i_H + g_{ij}\Omega^i_H\Omega^j_H=0$. 
In terms of matrix, it is
\begin{equation}
	\left(\begin{array}{ccc}
		1 ,& \Omega_{H}^{i}
	\end{array}\right) 
	\left(\begin{array}{ccc}
		g_{tt} & g_{jt}  \\
		g_{it} & g_{ij}
	\end{array}\right)_H
	\left(\begin{array}{ccc}
		1 \\ \Omega_{H}^{j}
	\end{array}\right) = 0 \,,
\end{equation}
where the lower subscrite $H$ represents taking value on the horizon. 
Notice that $g_{vv}$, $g_{iv}$ and $g_{ij}$ form a symmetric $(p+1)\times(p+1)$ matrix as the following 
\begin{equation} 
	K\equiv \left(\begin{array}{ccc}	
		g_{vv} & g_{jv} 
		\\		g_{iv} & g_{ij} 
	\end{array}\right)_H \;. \label{horizonKillingMatrix}
\end{equation}
Consequently, the column matrix $(1, \Omega_{H}^{i})^T$ should be the eigen vector of $K$ with zero eigen value:
\begin{equation}
	\left(\begin{array}{ccc}
		g_{tt} & g_{jt}  \\
		g_{it} & g_{ij}
	\end{array}\right)_H
	\left(\begin{array}{ccc}
		1 \\ \Omega_{H}^{j}
	\end{array}\right) 
	=  \left(\begin{array}{ccc}
		0 \\ 0
	\end{array}\right)  \,.
\end{equation}
It is equivalent to the following covariant relations:
\begin{equation} 
	(g_{\mu\nu} t^{\mu} \chi^{\nu})_H=(g_{\mu\nu} \psi_{i}^{\mu} \chi^{\nu})_H=0 \,.
\end{equation}
Therefore we obtain two conclusions: i) the determinant of $K$ is zero i.e., $\det(K)=0$; ii) the vectors $t^{\mu}$ and $\psi_{i}^{\mu}$ become spacelike and tangent to the horizon when locating on it.
For instanse, $\det{(K)}=0$ serves as $F(y^A)=0$ up to some factors. Its solution $y^A_H$ gives the location of the Killing horizon.

On the other hand, the horizon as a null hypersurface has a null normal covector $(\nabla^{\mu}F)_H$, such that
\begin{equation} 
	(g^{FF})_H \equiv (\nabla_{\mu}F\, \nabla^{\mu}F)_H =0\,. \label{gFFzero}
\end{equation}
It seems the situation of $\nabla_{\mu}F=0$ will ruin our argument, but $\nabla_{\mu}F=0$ also implies vanishing $g^{vF}$ and $g^{iF}$. We simply choose another function $\tilde{F}$ to ensure $\nabla_{\mu}\tilde{F}\neq 0$. Omitting the tilde sign, a nonvanishing $\nabla_{\mu}F$ at $F=0$ is null according to Eq.\eqref{gFFzero}. 
Since both of $(\chi^{\mu})_H$ and $\nabla^{\mu}F$ is the null normal vector of the horizon $F=0$, there should be should be the same with $(\chi^{\mu})_H=\alpha_H\nabla^{\mu}F$ on the horizon. Therefore, in the neighborhood the horizon, there is
\begin{equation}
	\chi^{\mu} = \alpha \nabla^{\mu} F +\mathcal{G}^{\mu} , \label{nearHext}
\end{equation}
where $\alpha$ is the analytic extension of $\alpha_H$ and $\mathcal{G}^{\mu}$ is a smooth vector field in the neighborhood of the horizon. The field $\mathcal{G}^{\mu}$ vanishes on the horizon, i.e., $(\mathcal{G}^{\mu})_H=0$.

\subsection{The proof}
Now we are well prepared to prove Eq.\eqref{sufgrapritrick}. 
At first, since the Killing vector field $\chi^{\mu}$ is $(\partial/\partial v)^{\mu}+\Omega^i_H (\partial/\partial \varphi^i)^{\mu}$, there are
\begin{equation}
	\chi^{\mu} \nabla_{\mu}v = 1 \,,\quad
	\chi^{\mu} \nabla_{\mu}\varphi^i  = \Omega^i_H \,,\quad
	\chi^{\mu} \nabla_{\mu}y^A   = 0 \, . \label{chiconcoor}
\end{equation}
We then constract Eq.\eqref{nearHext} with $\nabla_{\mu}v$, $\nabla_{\mu}\varphi$, and $\nabla_{\mu}y^A$, obtain
\begin{equation}
	\begin{split}
		1&=\alpha g^{vF}  +\mathcal{G}^{\mu}\nabla_{\mu}v \,, \\
		\Omega^i_{H}&=\alpha g^{iF}  +\mathcal{G}^{\mu}\nabla_{\mu}\varphi^i \,,\\
		0 &= \alpha g^{AF}  +\mathcal{G}^{\mu}\nabla_{\mu}y^A \,.
	\end{split}
\end{equation}
Notice that terms with $\mathcal{G}^{\mu}$ vanish and $\alpha$ becomes $\alpha_H$ if taking vale on the horiozn, we have 
\begin{equation}
	\begin{split}
		(g^{vF})_H &= (\nabla_{\mu}v\, \nabla^{\mu}F)_H = \frac{1}{\alpha_H} \,, \\
		(g^{iF})_H &= (\nabla_{\mu}\varphi^{i}\, \nabla^{\mu}F)_H = \frac{\Omega_{H}^{i}}{\alpha_H} \,,\\
		(g^{AF})_H &= (\nabla_{\mu}y^A\, \nabla^{\mu}F)_H =0 \,.
	\end{split}
\end{equation}
Hence we obtain $\Omega^i_H = (g^{iF}/g^{vF})_H$, the second equation in
Eq.\eqref{sufgrapritrick}.

Next, we will calculate the surface gravity $\kappa$.
Applying the Killing equation $\nabla_{\mu}\chi_{\nu}+\nabla_{\nu}\chi_{\mu}=0$ leads to
\begin{equation}
	\chi^{\nu}\nabla_{\nu}\chi_{\mu} =-\chi^{\nu}\nabla_{\mu}\chi_{\nu}=-\frac{1}{2}\nabla_{\mu} (\chi_{\nu}\chi^{\nu}) \,. \label{changemunuorder}
\end{equation}
Such a treatment hints at claculating $\chi_{\nu}\chi^{\nu}$. Using Eq.\eqref{nearHext}, we have
\begin{equation}
	\begin{split}
		\chi^{\mu}\chi_{\mu}=\alpha^2 g^{FF} +2\alpha \mathcal{G}^{\mu}\nabla_{\mu}F +\mathcal{G}^{\mu} \mathcal{G}_{\mu}\,. 
	\end{split}
\end{equation}
Since $\chi^{\mu}\nabla_{\mu}y^A=0$ and $F(y^A)$ does not depends on $v$ and $\varphi^2$, we have $\chi^{\mu}\nabla_{\mu}F=\alpha g^{FF}+ \mathcal{G}^{\mu} \nabla_{\mu} F=0$. Therefore,  
\begin{equation}
	\begin{split}
		\chi^{\mu}\chi_{\mu}=-\alpha^2 g^{FF}+\mathcal{G}^{\mu} \mathcal{G}_{\mu} \,.
	\end{split}
\end{equation}
Directly derivate the above result for $\chi^{\mu}\chi_{\mu}$, we obtain
\begin{equation}
	\chi^{\nu}\nabla_{\mu}\chi_{\nu}
	=-\frac{\alpha^2}{2} \nabla_{\mu}g^{FF}
	-\frac{g^{FF}}{2} \nabla_{\mu}\alpha^2
	+\mathcal{G}^{\nu} \nabla_{\mu}\mathcal{G}_{\nu}  \,. \label{chiandgragFF}
\end{equation}
The second term and the third term would vanish if we let it take value on the horizon. For instanse, Eq.\eqref{surfacefravitydef}, Eq.\eqref{changemunuorder} and Eq.\eqref{chiandgragFF} leads to
\begin{equation}
	\kappa(\chi^{\mu})_H=\frac{1}{2(g^{vF})^2_H} (\nabla^{\mu}g^{FF})_H \,,
\end{equation}
where we replace $\alpha_H$ by $1/(g^{vF})_H$.
Apply the contraction $\chi^{\mu}\nabla_{\mu}v=1$ again, the formula for $\kappa$ given in Eq.\eqref{sufgrapritrick} is confirmed to complete the proof.

\section{Examples} 
This section will give three examples that are concrete but still 
general to show how to apply Eq.\eqref{sufgrapritrick}.
Particularly, when dealing with a rotating black hole, focusing on the inverse metric takes more advantages, like easily identifying the pole indicating the Killing horizon location, and directly reading relevant components to calculate surface gravities and angular velocities.  
%Issue about choosing nice slices is also included.

\subsection{Static spherical spacetime}  
The metric of a general static spherical spacetime is
\begin{equation}
	ds^2=-e^{-2\chi(r)} f(r) dt^2 +\frac{dr^2}{f(r)} +r^2d\Omega^2  \,, %(d\theta^2+\sin^2\theta d\phi^2)
\end{equation}
under the orthogonal coordinates. The $t$-$r$ part hints at taking out the following structure  
\begin{equation}
	-e^{-2\chi} f(dt^2- (e^{\chi} \frac{dr}{f})^2)=-e^{-2\chi} f(dt+e^{\chi} \frac{dr}{f})(dt-e^{\chi} \frac{dr}{f}) \,,
\end{equation}
where the argument $r$ of functions $f$ and $\chi$ is omitted.
Usually, the \textit{retard} Eddington-Finkelstein (EF) time is $du=dt-e^{\chi}dr/f$ while the \textit{advanced} EF time is $dv=dt+e^{\chi}dr/f$. Here we only use the notation $v$ to write down the expression in a unified way. Hence the EF time is written as $dv=dt\pm e^{\chi}dr/f$ such that the metric in the EF coordinates is 
\begin{equation}
	ds^2=-e^{-2\chi} f dv^2 \pm 2e^{-\chi}dvdr +r^2 d\Omega^2 \,.
\end{equation}
On the other hand, it is easier to obtain components of the inverse metric via coordinates transformation. Considering we only change the time coordinate, components of the inverse metric related to $v$ may be affected. A direct calculation shows that
\begin{equation}
	g^{vv}=g^{tt}+\frac{e^{2\chi}}{f^2}g^{rr}=0 \,,\quad
	g^{vr}=\pm\frac{e^{\chi}}{f}g^{rr}=\pm e^{\chi} \,,\quad
	g^{v\theta}=g^{v\phi}=0 \,,
\end{equation}
while other components remain unchanged, i.e., $g^{rr}=f$, $g^{\theta\theta}=1/r^2$ and $g^{\phi\phi}=1/(r^2\sin^2\theta)$. 
These are all components of inverse metric under the coordinates $\{v,r,\theta, \phi\}$.

The relevant Killing vector is $\chi^{\mu}(\partial/\partial x^{\mu})=\partial/\partial t=\partial/\partial v$ while coordinates $\{r,\theta, \phi\}$ are the Frobenius scalars. It is the case of $p=0$. %and $n=4$. 
Furthermore, the root $r=r_H$ of equation $f(r)=0$ determines the horizon location if $\chi(r)$ is regular. Usually, the function $\chi(r)$ is $0$, like the famous Shcwarzchild black hole and Reissner-Nordstr\"om (RN) black hole. Excluding the situation of blowing up $\chi(r_H)$, $f(r_H)=0$ indicates that $r=r_H$ is a null hypersurface ($(g^{rr})_H=0$ without $ (\nabla_{\mu}r)_H=0$). The Killing vector $\chi^{\mu}$ becomes null there so $r=r_H$ describes a Killing horizon. Applying Eq.\eqref{sufgrapritrick}, we obtain
\begin{equation}
	\kappa=\pm \frac{1}{2}e^{-\chi_H} f'_H \,.\label{spkappa}
\end{equation}
There is no angular momentum because we are considering spacetimes with spherical symmetry. In addition, applying PG coordinates or other time coordinate labeling nice slices leads to the same result for $\kappa$. The key feature is that the pole of $g^{tt}$ due to $1/f$ is canceled out. Such a canceling implies the same non-zero value of $(g^{\tilde{t} r})_H$ in which $\tilde{t}$ labels another set of nice slices.

It is worth noting that $\kappa$ may be negative. If $f'(r_H)>0$, the positive $f$ region is outside of the negative $f$ region, i.e., $f>0$ in  $r>r_H$. Applying EF coordinates, it is the set of advanced slices 
that penetrates the black hole horizon. Thus Eq.\eqref{spkappa} implies a positive surface gravity on the black hole horizon. While the retarded EF coordinates response to a white hole horizon with a negative surface gravity. 
On the other hand, suppose $f$ becomes zero at $r=r_C$ and satisfies $f'(r_C)<0$, then $\kappa$ is positive in the retarded EF slices (taking $-$ sign); negative the advanced EF slices (taking $+$ sign). For instance, the inner Cauchy horizon in RN spacetime and the cosmic horizon in Schwarzchild-de Sitter spacetime are cases of $f'(r_C)<0$.
%In conclusion, the $\kappa$ is positive for the black hole horizon and the cosmic horizon of an expanding universe; while it is negative if taking the case of the white hole or the shrinking universe.
It is recommended to read Refs.\cite{WaldGR, LiangGRvIII} to understand the sign of $\kappa$ on four types of non-extreme horizons. Killing horizons are also trapping horizons, thus the issue of types also appears when studying non-degenerate trapping horizons, see Refs.\cite{Hayward:1993wb, Hayward:1994bu, Hayward:1997jp, Hayward:1998pp, Hayward:2005gi, Helou:2015zma, Helou:2016xyu, Binetruy:2018jfz, Yang:2021diz}.

%The topic about embedding a black hole in the 
%The Schwarzchild-de Sitter spacetime is a good example to discuss changing slices.
%\cite{}
%\begin{equation}
%    \chi=0 \,,\quad
%	f=1- \frac{2GM}{r} + \frac{\Lambda\,r^2}{3} \,,
%\end{equation}

%To handle the locations of the black hole horizon and the cosmic horizon, we choose

%Such a classification is consistent with what Refs.\cite{} do for the so-called trapping horizon. We prefer to use terms in a unified way. Hence 

\subsection{Generalized Kerr-Newnman metric}
%	We use the KN black hole to show how to use the above tools.
Ref.\cite{Baines:2023dhq} proposed an interesting metric generalizing the Kerr-Newnman(KN) metric. It takes the following form:
\begin{equation}
	\begin{split}
		ds^2
		=&~-\frac{\Delta e^{-2\Phi} -a^2\sin^2\theta}{\Xi^2+a^2\cos^2\theta} dt^2 
		+\frac{\Xi^2+a^2\cos^2\theta}{\Delta} dr^2 +(\Xi^2+a^2\cos^2\theta) d\theta^2
		\\&-2 \frac{a\sin^2\theta(\Xi^2-\Delta e^{-2\Phi}+a^2 ) }{\Xi^2+a^2\cos^2\theta} dtd\phi
		+ \frac{ (\Xi^2+a^2)^2 -2e^{-2\Phi} \Delta a^2\sin^2\theta}{\Xi^2+a^2\cos^2\theta}\sin^2\theta d\phi^2  \,, \label{BVKerr}
	\end{split}
\end{equation}
in which $\Delta$, $\Xi$, and $\Phi$ are functions of $r$ and we have omitted the argument $r$ for short. The metric contains non-vanishing $g_{tt}$, $g_{t\phi}$, $g_{\phi\phi}$, $g_{rr}$ and $g_{\theta\theta}$ terms. 
It reduces to the Kerr spacetime when taking $\Phi=0$, $\Xi=r$, and $\Delta=r^2+a^2-2Mr$ (in the geometric units).

We need to find out the inverse metric to apply our trick.
It would be beneficial to calculate $g_{tt}g_{\phi\phi}-g_{t\phi}^2$ first:
\begin{equation}
	g_{tt}g_{\phi\phi}-g_{t\phi}^2= -\Delta e^{-2\Phi} \sin^2\theta \,.
\end{equation}
%The first step is to find out the inverse metric components.
Then it helps us to calculate the inverse metric 
\begin{equation}
	\begin{split}
		g^{tt}=&~ -\frac{e^{2\Phi}(\Xi^2+a^2)^2}{\Delta (\Xi^2+a^2\cos^2\theta)} + \frac{a^2\sin^2\theta}{\Xi^2+a^2\cos^2\theta} \,, \quad
		\\	g^{t\phi}=&~ - \frac{a e^{2\Phi}(\Xi^2+a^2)}{\Delta (\Xi^2+a^2\cos^2\theta)} + \frac{a}{\Xi^2+a^2\cos^2\theta}  \,, \quad
		\\	g^{\phi\phi}=&~ - \frac{a^2e^{2\Phi}}{\Delta (\Xi^2+a^2\cos^2\theta)} + \frac{\csc^2\theta}{\Xi^2+a^2\cos^2\theta} \,,
		\\ g^{rr}=&~ \frac{\Delta}{\Xi^2+a^2\cos^2\theta} \,,\qquad\qquad
		g^{\theta\theta}=\frac{1}{\Xi^2+a^2\cos^2\theta} \,. \label{ing3funBM}
	\end{split}
\end{equation}
The inverse metric hints at a suitable coordinates transformation. The equation $\Delta=0$ leads to $g^{rr}=0$ such that the root of $\Delta(r)=0$ determines the location of the Killing horizon. One can choose $\Delta(r)=0$ as $F=0$, but the choice $F=r-r_H=0$ will be more convenient. Nevertheless, there is $\nabla_{\mu}\Delta=\Delta'\nabla_{\mu}r$. In order to cancel the pole $\Delta=0$ in Eq.\eqref{ing3funBM}, we make the following ansatz for the coordinates transformation:
\begin{equation}
	\begin{split}
		dv= dt +\frac{P(r)}{\Delta(r)}dr \,,\quad  d\varphi= d\phi +\frac{\omega(r)P(r)}{\Delta(r)}dr \,. \label{prepareEF}
	\end{split}
\end{equation}
Therefore, relevant components of the inverse metric in the new coordinates are
\begin{equation}
	\begin{split}
		g^{vv}&%= g^{tt} +\frac{P^2}{\Delta (a^2\cos^2\theta+\Xi^2)} 
		=\frac{P^2-e^{2\Phi}(\Xi^2+a^2)^2}{\Delta (\Xi^2+a^2\cos^2\theta)} + \frac{a^2\sin^2\theta}{\Xi^2+a^2\cos^2\theta}
		\,,\quad  
		\\	g^{v\varphi}&%= g^{t\phi} +\frac{\omega P^2}{\Delta (a^2\cos^2\theta+\Xi^2)}
		= \frac{\omega P^2-a e^{2\Phi}(\Xi^2+a^2)}{\Delta (\Xi^2+a^2\cos^2\theta)} + \frac{a}{\Xi^2+a^2\cos^2\theta}  \,,\quad
		\\	g^{\varphi\varphi}&%= g^{\phi\phi} +\frac{\omega^2 P^2}{\Delta (a^2\cos^2\theta+\Xi^2)} 
		=\frac{\omega^2P^2-a^2e^{2\Phi}}{\Delta (\Xi^2+a^2\cos^2\theta)} + \frac{\csc^2\theta}{\Xi^2+a^2\cos^2\theta} \,.
	\end{split}
\end{equation}
The simplest option to cancel the pole is to require terms with pole $\Delta=0$ disappear. It leads to
\begin{equation}
	P =\pm e^{\Phi}(\Xi^2+a^2) \,,\quad  \omega=\frac{a}{a^2+\Xi^2}  \,, %\omega(r_H) =
\end{equation}
where the $+$ sign is for the generalized advanced Eddington-Finkelstein(EF) coordinates while the $-$ sign is for the generalized retarded EF coordinates. 
Then relevant inverse metric components take values as
\begin{equation}
	(g^{vr})_H=\pm e^{\Phi_H}\frac{\Xi_H^2+a^2}{a^2\cos^2\theta+\Xi_H^2}  \,,\quad
	(g^{\varphi r})_H=\pm e^{\Phi_H}\frac{a}{a^2\cos^2\theta+\Xi_H^2}  \,,
\end{equation}
on the horizon. 
Such values are not affected by another choice for $P$, like $P=\pm e^{\Phi}(\Xi^2+a^2)+\Delta(r)W(r)$ in which $W(r)$ is any smooth function of $r$ at $r=r_H$. Some generalized PG coordinates for the KN black hole are proposed in Refs.\cite{Doran:1999gb, Jiang:2005ba, Hobson:2022cso}. They are specific choices of $W(r)$, while the generalized EF coordinate can be viewed as taking $W(r)=0$. Other generalized PG coordinates for KN black hole in Refs.\cite{Zhang:2005gja, Aly:2023qzk} also give nice slices.  
Nevertheless, the derivative of $g^{rr}$ on the horizon is
\begin{equation}
	(\nabla_{\mu}g^{rr})_H= \frac{\Delta'_H}{\Xi_H+a^2\cos^2\theta}(\nabla_{\mu}r)_H\,. \label{gradgrr}
\end{equation}
Therefore, applying Eq.\eqref{sufgrapritrick}, we obtain
\begin{equation}
	\kappa=\pm \frac{ e^{-\Phi_H} \Delta'_H}{2 (\Xi_H^2+a^2)} \,,\quad 
	\Omega_H= \frac{a}{a^2+\Xi_H^2} \,.
\end{equation}
The surface gravity matches the remarkably simple formula given by Refs.\cite{Baines:2023dhq, Baines:2023cac}. 
Furthermore, the issue for the sign of $\kappa$ appears again. Taking Kerr-Newnman metric ($\Xi=r,\,\Phi=0,\,\Delta=r^2-2Mr+a^2+Q^2$) as an example, $\Delta=0$ contains two roots: $r_{\pm}=M\pm\sqrt{M^2-a^2-Q^2}$ and the derivative of $\Delta$ there takes $\Delta'(r_+)>0$ and $\Delta'(r_-)<0$. 
Meanwhile, choosing $P =r^2+a^2$ leads to the \text{advanced} time slices which penetrate the black hole horizon $r_+$ and the future Cauchy horizon $r_-$. Then the surface gravity is positive at $r=r_+$ and negative at $r=r_-$. Choosing $P =-(r^2+a^2)$ implies opposite results.

%Such a treatment gives a systematic way to generalize the Kerr metric. The crucial is to ensure the pole is canceled.
%Notice that the sign

%\input{HJeq.tex}
\subsection{Generalized Kerr-de Sitter metric}
The metric Eq.\eqref{BVKerr} does not cover the case of Kerr-de Sitter(Kerr-dS) spacetime.
We thus propose a four-functions metric to generalize the Kerr-dS metric. 
\begin{equation}
	\begin{split}
		ds^2=&~ -\frac{\Delta_r\,e^{-2\Phi}}{\Xi^2+a^2\cos^2\theta}\big( dt- \frac{a\sin^2\theta}{1+\frac{\Lambda\,a^2}{3}}d\phi \big)^2 +(\Xi^2+a^2\cos^2\theta)\big( \frac{dr^2 }{\Delta_r}  + \frac{d\theta^2}{\Delta_\theta}  \big) \\&+\frac{\Delta_\theta\sin^2\theta}{\Xi^2+a^2\cos^2\theta}\big( -a dt+ \frac{\Xi^2+a^2}{1+\frac{\Lambda\,a^2}{3}} d\phi\big)^2 \,,\label{genKerrdS}
	\end{split}
\end{equation}
where $\Phi,\Xi,\Delta_r$ are functions of $r$ and $\Delta_\theta$ is a function of $\theta$.
The Kerr-de Sitter metric is a specific situation when
\begin{equation}
	\Phi=0\,,\;\Xi=r \,,\;
	\Delta_r=(r^2+ a^2)(1-\frac{\Lambda r^2}{3})-2M r \,,\;
	\Delta_\theta=1+\frac{\Lambda a^2}{3}\cos^2\theta \,. \label{KdS2Delta}
\end{equation}
Hence the four-function metric Eq.\eqref{genKerrdS} can include situations with nonzero cosmological constant. It may be beneficial to study a black hole embedded in an exponentially expanding universe.
Its inverse metric has the following non-zero components:
\begin{equation}
	\begin{split}
		g^{tt}=&~ -\frac{e^{2\Phi}(\Xi^2+a^2)^2}{\Delta_r (\Xi^2+a^2\cos^2\theta)} + \frac{a^2\sin^2\theta}{\Delta_\theta (\Xi^2+a^2\cos^2\theta)} \,, \quad
		\\	g^{t\phi}=&~(1+\frac{\Lambda}{3} a^2) (- \frac{a e^{2\Phi}(\Xi^2+a^2)}{\Delta_r (\Xi^2+a^2\cos^2\theta)} + \frac{a}{\Delta_\theta (\Xi^2+a^2\cos^2\theta)} ) \,, \quad
		\\	g^{\phi\phi}=&~(1+\frac{\Lambda}{3} a^2)^2 ( - \frac{a^2e^{2\Phi}}{\Delta_r (\Xi^2+a^2\cos^2\theta)} + \frac{\csc^2\theta}{\Delta_\theta (\Xi^2+a^2\cos^2\theta)}) \,,
		\\ g^{rr}=&~ \frac{\Delta_r}{\Xi^2+a^2\cos^2\theta} \,,\qquad\qquad
		g^{\theta\theta}=\frac{\Delta_\theta}{\Xi^2+a^2\cos^2\theta} \,. \label{ing4fungKdS}
	\end{split}
\end{equation}
Such an inverse metric also indicates that the Hamilton-Jacobi equation for geodesics $g^{\mu\nu}\nabla_{\mu}S\nabla_{\nu}S=-\mu^2$ is separable when considering $S=-\omega t+ m\phi+\mathcal{N}(r)+\Theta(\theta)$. Hence it may be interesting to study the image of the metric Eq.\eqref{genKerrdS} like those research in Refs.\cite{,}.
Moreover, a coordinate transformation similar to Eq.\eqref{prepareEF} leads to regular coordinates on the horizon. Concretely, taking
\begin{equation}
	\begin{split}
		dv= dt \pm e^{\Phi}(\Xi^2+a^2) \frac{dr}{\Delta_r} \,,\quad  d\varphi= d\phi \pm e^{\Phi}a(1+\frac{\Lambda}{3} a^2) \frac{dr}{\Delta_r}\,. \label{EFforgenKdS}
	\end{split}
\end{equation}
implies that
\begin{equation}
	\begin{split}
		g^{vv}&= \frac{a^2\sin^2\theta}{\Delta_\theta (\Xi^2+a^2\cos^2\theta)}
		\,,\;  
		g^{v\varphi}= \frac{a(1+\Lambda a^2/3)}{\Delta_\theta (\Xi^2+a^2\cos^2\theta)}    \,,\;
		g^{\varphi\varphi}=  \frac{(1+\Lambda a^2/3)^2\csc^2\theta}{\Delta_\theta (\Xi^2+a^2\cos^2\theta)} \,, \\
		g^{vr}&=\pm  \frac{e^{\Phi}(\Xi^2+a^2)}{\Xi^2+a^2\cos^2\theta} 	\,,\qquad  
		g^{\varphi r}=\pm (1+\frac{\Lambda}{3} a^2) \frac{e^{\Phi}a}{\Xi^2+a^2\cos^2\theta}\,,
	\end{split}
\end{equation}
Since $g^{rr}$ still takes the expression as $\Delta_r/(\Xi^2+a^2\cos^2\theta)$, its derivative is $(\Delta'_r\nabla_{\mu}r)_H/(\Xi_H^2+a^2\cos^2\theta)$ at $r=r_H$ which is the root of $\Delta_r=0$ . Therefore, Eq.\eqref{sufgrapritrick} implies that
\begin{equation}
	\kappa=\pm \frac{ e^{-\Phi_H} \Delta'_{rH}}{2 (\Xi_H^2+a^2)} \,,\quad 
	\Omega_H=(1+\frac{\Lambda}{3} a^2) \frac{a}{a^2+\Xi_H^2} \,.
\end{equation}
For the Kerr-dS ($\Lambda>0$), the largest root of equation $\Delta_r=(r^2+ a^2)(1-\Lambda r^2/3)-2M r=0$ indicates the location of the cosmic horizon; the second large one is the black/white hole horizon while the smallest positive root corresponds the inner Cauchy horizon.

\section{Summary}
The theorem proved in section 2 leads to a trick for calculating the Killing horizon surface gravity $\kappa$. 
Once the coordinates frame with nice slices regularly penetrating the Killing horizon is chosen, $\kappa$ can be calculated via particular inverse metric components under such coordinates by Eq.\eqref{sufgrapritrick}, with no need to calculate Christoffel symbols.

We take spherically symmetric spacetime, the three-function generalized Kerr metric proposed in Ref.\cite{Baines:2023dhq}, and the four-function generalized Kerr-dS metric as examples to show detailed steps of the trick.
In practice, starting from a stationary metric under the orthogonal coordinates, the inverse metric shows poles that indicate Killing horizons in relevant components. Those poles hint at suitable coordinates transformations for choosing nice slices. Then the inverse metric components under the new coordinates frame directly lead to the surface gravity and angular velocities through Eq.\eqref{sufgrapritrick}. Moreover, the trick is also beneficial to handle the sign of the surface gravity which distinguishes four types of non-degenerate horizons. Therefore, not limited to situations of the black hole or white hole, cases like the inner Cauchy horizon and the cosmic horizon are also covered.

\bibliographystyle{JHEP}
\bibliography{RefSurfaceGravity.bib} 
\end{CJK}
\end{document}